# Magic-angle bilayer graphene nano-calorimeters - towards broadband, energy-resolving single photon detection

P. Seifert[1], X. Lu[1], P. Stepanov[1], J. R. Duran[1], J. N. Moore[1], K. C. Fong[2,3], A. Principi[4] and D. K. Efetov[1*]

1. ICFO - Institut de Ciencies Fotoniques, The Barcelona Institute of Science and Technology, Castelldefels, Barcelona, 08860, Spain
2. Raytheon BBN Technologies, Quantum Information Processing Group, Cambridge, Massachusetts 02138, USA
3. Department of Physics, Harvard University, Cambridge, Massachusetts 02138, USA
4. Department of Physics and Astronomy, The University of Manchester, Oxford Road, M13 9PL, Manchester (UK)

*E-mail: dmitri.efetov@icfo.eu

**Abstract:**
**Because of the ultra-low photon energies in the mid-infrared and terahertz frequencies, in these bands photodetectors are notoriously underdeveloped, and broadband single photon detectors (SPDs) are non-existent. Advanced SPDs exploit thermal effects in nano-structured superconductors, and their performance is currently limited to the more energetic near-infrared photons due to their high electronic heat capacity. Here, we demonstrate a superconducting magic-angle twisted bilayer graphene (MAG) device that is capable of detecting single photons of ultra-low energies by utilizing its record-low heat capacity and sharp superconducting transition. We theoretically quantify its calorimetric photoresponse and estimate its detection limits. This device allows the detection of ultra-broad range single photons from the visible to sub-THz with response time around 4 ns and energy resolution better than 1 THz. These attributes position MAG as an exceptional material for long-wavelength single photon sensing, which could revolutionize such disparate fields as quantum information processing and radio astronomy.**

**Introduction:**
The detection of single photons is a key enabling technology in many research areas including quantum sensing, quantum key distribution, information processing and radio astronomy. Single photon detectors (SPDs) for wavelengths ranging from the visible to near infrared (nIR) have already been developed and even commercialized. State-of-the-art SPD technologies rely on heat-induced breaking of the superconducting state in nano-structured superconductors (SCs). Here, superconducting transition-edge sensors (TES) and superconducting nanowire single photon detectors (SNSPDs) have developed into the SPDs with the highest detection efficiencies, lowest dark count rates and operation wavelengths up to 10 μm (*1–8*). However, while no theoretical performance limits are evident so far, extending the broadband detection of single photons from the nIR to the far-infrared and the terahertz (THz), has yet to be demonstrated.

TESs exploit the steepness of the temperature dependent resistance at the superconducting transition edge, which enables the generation of detectable voltages pulses upon heating electrons by absorbed light quanta(*4, 5, 9*). Because the energy of an absorbed photon is transferred to the whole ensemble of electrons, the performance of TESs is determined by the heat capacity of the calorimetric materials used. This currently limits the SPD operation of TESs to wavelengths below 8μm(*10*), temperatures below 100mK, and detection times above ~10μs(*10–13*).

Strategies to reduce the heat capacity have led to a targeted development of ever-thinner nanostructured SC thin films and the use of low carrier density SCs so that absorbed heat is shared among fewer electrons. However, traditional material fabrication approaches have set these developments a limit. The SC thin films are strongly disordered, polycrystalline, and have thicknesses exceeding several nanometres because they are obtained from high-electron-density SCs by sputtering and etching.

Preparing SPDs from the newly discovered two-dimensional (2D) superconductor magic-angle twisted bilayer graphene (MAG, Fig. 1 a) has the potential to dramatically break through these limits. By stacking two graphene layers on top of each other with a relative twist angle between them, a "moiré" pattern gives rise to a long-wavelength periodic potential(*14*). It was shown that for a well-defined twist angle of 1.1°, the so-called "magic" angle, flat bands with ultra-high density of states are formed, and interactions give rise to correlated insulating and dome-shaped superconducting phases with a $T_c$ > 3K(*15–17*). Figure 1 (b, inset) shows an experimentally obtained phase diagram of our MAG device. .Resistance $R$ is plotted as a function of carrier density $n$ and temperature $T$. Here the superconducting phase occurs at a record-low electron density, as summarized in Figure 1 (b). This density is four orders of magnitude lower than in traditional SC thin films used in SPD applications, and at least an order of magnitude lower than in other 2D van der Waals SCs. These attributes, together with its thickness of only 0.6nm and its ultra-high crystallographic quality, make MAG a promising candidate to extend SPD operation to previously unimaginable wavelengths in the mid-IR and even THz range.

**Results and discussion:**
In this manuscript, we explore the feasibility of preparing energy-resolved SPDs from MAG, by estimating its thermal response due to the absorption of single photons. Due to the steep temperature dependent resistance at its SC transition edge, photon-generated voltage pulses can be directly read out. Figure 1 (a) depicts the schematics of the proposed detector, comprised of an electrically contacted 250nm x 250nm superconducting MAG sheet on top of a Si/SiO$_2$ substrate, which acts as a capacitive back gate for tuning $n$. As in conventional SCs, the MAG´s superconducting gap vanishes at the critical temperature, allowing in principle a broadband optical absorption. The challenge becomes the relatively low absorption coefficient of graphene (~2.3%) as well as a small detector area. Fortunately, several approaches have been successfully developed and implemented to enhance the absorption of graphene to almost 100% by coupling it with photonic crystals(*18–20*), Fabry-Pérot microcavities(*21*) or ring resonators(*22*) for operation at nIR and mid-IR wavelengths, as well as integrations with a planar antenna for THz operation(*23*). Ultimately, optical coupling approaches will be determined by the demands of the anticipated application, as no single device architecture can enhance it in the broad range from the THz the near-IR.

After an electron absorbs a photon, the energy is thermalized within the electron bath via electron-electron scattering on a timescale of ~100fs, which is much shorter than the relaxation time to thermal equilibrium(*24*). Hence, photo-excited electrons can be described by a Fermi-Dirac distribution characterized by an effective (electronic) temperature which is different from that of the lattice(*24*). We start by quantifying the thermal properties of MAG by calculating its temperature dependent electronic heat capacity $C_e(T)$. Using the continuum model described in Ref. (*25*), we first calculate the single-particle band structure of MAG. As known, ultra-flat bands occur close to charge neutrality (red bands in Fig. 2 a, inset). We then extract the density of states of the moiré bands and from it calculate $C_e(T)$ (see supplementary note 1). Figure 2 (a) shows $C_e(T)$ as a function of temperature $T$ for $n = 1.1 \cdot 10^{12}/\text{cm}^2$, which coincides with the density of one SC dome. $C_e(T)$ does not exhibit a feature at the SC phase transition because no microscopic model of interactions was applied in this calculation. Remarkably, the values of the heat capacity at temperatures close to the SC $T_c$ are extremely small, of the order of a few hundred $k_B$, which is 2-3 orders of magnitude lower than in any other superconducting single photon detector (exhibiting lowest heat capacities in the range of $10^4$-$10^5 \, k_B$ (26)).

By equating the energy of an incident photon with the absorption-induced increase in internal energy $\hbar\omega = \int_{T_0}^{T_{max}} C_e(T)dT$, we can now calculate the temperature increase $\Delta T$ of the electrons in the MAG sheet upon photon absorption. Figure 2 (b) depicts the corresponding $\Delta T$ as a function of photon frequency for different temperatures $T$. Remarkably, for $T \lesssim T_c$, we find relatively large values for $\Delta T$ on the order of several kelvin for the absorption of mid-IR photons; and even for photons in the THz and 100-GHz frequencies, $\Delta T$ remains sizable, on the order of 10-100mK. To achieve an optimum detection performance, a sharp SC transition is highly desirable, as it enables detectable voltage pulses to be generated even from weak photon-induced heat pulses. Figure 2 (c) shows experimentally obtained $R$ and $dR/dT$ as a function of temperature $T$ for optimal doping of the superconducting dome at $n = 1.1 \cdot 10^{12}/\text{cm}^2$. Around a critical temperature of $T_c = 0.65$ K, the device exhibits a very sharp transition edge with a large resistance change.

In order to evaluate the intrinsic detector performance, we can extract the photon-induced voltage change $\Delta V$ across a current-biased MAG sheet (see supplementary note 2). This can be achieved by combining the temperature dependent resistivity at the superconducting transition with the calculated temperature change $\Delta T$ due to the absorption of a single photon. Figure 3 (a) inset shows the experimentally obtained $I/V$ characteristics of the MAG device, where we find a superconducting critical current $I_c$ > 20nA. To maximize $\Delta V$, the device is current biased just below $I_c$ to $I$ = 20nA. The so-obtained $\Delta V$ is shown in Figure 3 (a) as a function of $T$ and frequency of the absorbed photon $f_p$. Strikingly, we find relatively large voltage signals on the order of tens of μV for a very broad range of photon frequencies, from the nIR all the way to ~100GHz.

The lifetime of the voltage pulses is determined by the intrinsic thermal relaxation pathways of the thermally excited electrons in the MAG sheet. Here, as is well established for single-layer graphene devices(27), we assume that the dominant heat dissipation channels are 1) electron interaction with acoustic phonons ($G_{e-ph}$) and 2) heat diffusion to the electrodes by the Wiedemann-Franz law ($G_{WF}$) (27). The corresponding thermal conductivities are plotted in Figure 3 (b). In contrast to single-layer graphene, where the electron-phonon interaction at low $T$ is relatively small, we find that in MAG $G_{e-ph}$ dominates over $G_{WF}$ by several orders of magnitude at all temperatures. We do not consider cooling via optical phonon scattering because the energies of hot electrons here are much smaller than those of optical phonons. We also ignore radiative cooling ($G_{rad}$) because $G_{rad}$ can be estimated to $\sim k_B \cdot B$, where $B$ is the measurement bandwidth, and this yields values which are at least 5 orders of magnitude smaller than $G_{e-ph}$ for $B$ = 1 GHz (28, 29).

Having established $G_{e-ph}$ as the dominant heat relaxation mechanism, we obtain the thermal relaxation time $\tau$ for different $T$ through the quasi-equilibrium relation $G_{e-ph} \cdot \tau = C_e$, as is shown in Figure 3 (b) (see supplementary note 1). Figure 3 (c) shows the transient thermal response of the device after photon absorption at a temperature of $T = T_c$ for photon frequencies between 0.5 THz and 20 THz (see also supplementary figures 1 and 2). Remarkably, at $T = T_c$ the hot electron distribution relaxes within ~4 ns for all photon frequencies. This is orders of magnitude faster than other energy-resolving superconducting single photon detectors, which exhibit recovery times on the order of tens of microseconds (e.g. in TESs(5)). We note that at lower device temperatures the decreasing electron-phonon interaction leads to strongly increasing relaxation times surpassing 100 ns at $T = 0.3$ K and even 1 μs at $T = 0.25$ K (compare supplementary figure 1 (a)).

Depending on the final detector architecture, the fast intrinsic photo-response of the MAG can be further processed with broadband low-noise amplifiers, such as HEMT. In the future, we can also employ a frequency multiplexed readout, similar to KID, based on the change of the kinetic inductance of MAG in a resonator(*30*). Not only does this allows MAG detector arrays, but it also leads to even higher sensitivities by virtue of lower temperature operation.

It is also possible to resolve the energy of an absorbed photon from the detector's transient voltage response because the amplitude of that response increases monotonically with increasing photon frequency. At a bandwidth on the order of $1/\tau$, the energy resolution of a calorimeter without feedback is limited by thermodynamic energy fluctuations $\langle \Delta E^2 \rangle = k_B T^2 C_e$ (*31*). One can understand these thermodynamic fluctuations in terms of random fluctuations of the internal energy of the electron distribution due to its statistical nature as a canonical ensemble in thermal exchange with the bath(*31*). For a given energy $E$ of an absorbed photon, these energetic fluctuations cause an uncertainty in the energy measurement which results in a Gaussian distribution of potentially measured energies around $E$ with width $\Delta E$. While for high energy photons downconversion mechanisms involving optical phonons can further limit the energy resolution(*32*), in our detectors these processes can be largely neglected. This is due to the very low energies of THz photons, and the combination of graphene's ultra-high optical photon energies and its weak electron-phonon interaction.

Figure 4 (a) illustrates the measurement uncertainty with an exemplary histogram of the energy distribution that would be generated by detecting one thousand photons at two different energies. Due to energy broadening, two photon energies can be distinguished only if their energies are separated by more than the full width at half maximum of the distribution in $E$, equal to $\sim 2.35 \Delta E$. Figure 4 (b) shows the energy scale $\Delta E = \sqrt{\langle \Delta E^2 \rangle}$ (red) and the associated relative temperature $\delta T = \Delta E / C_e$ (blue) of these fluctuations as a function of the device temperature. At $T_c$, the energy fluctuations are on the order of $\Delta E \sim 1$ meV and the relative temperature fluctuations are on the order of $\delta T < 0.1\, T$. Figure 4 (c) shows the thermodynamic fluctuations limited frequency resolution $\Delta f_{\text{ph}} = 2.35 \cdot \frac{\Delta E}{h}$. Before photon absorption (black dashed line), we find that our device allows for an energy resolution better than 1 THz, which is on par with the most sensitive calorimeters for THz applications (*26*). Given the large increase in electron temperature in response to a single photon (Figure 2 (b)) and the strong temperature dependence of $\langle \Delta E^2 \rangle$, the energy resolution will be inevitably reduced upon absorption of a photon. We assume a frequency dependent photon induced effective temperature increase $T_{\text{eff}}(f_p)$ to modify $\langle \Delta E^2 \rangle (f_p)$ according to (*33*). The solid lines in Figure 4 (c, d) show the corresponding frequency resolution for the absorption of a photon as a function of temperature and photon frequency. While at frequencies higher than 10 THz we see a significant decrease in frequency resolution, the resolution in the few-THz regime is not strongly affected and stays below 1 THz. Importantly, the relative frequency resolution is always $\Delta f / f < 1$ (outside the shaded area in Figure 4 (d)), allowing frequency resolved single photon detection.

**Conclusion:**
In conclusion, we demonstrate a superconducting magic-angle twisted bilayer graphene device that has a very high potential application as single photon detector. The two-dimensional, highly crystaline nature of the bilayer graphene enables an easy integration of the device into THz antennas or microwave resonators. By exploring the device characteristics and theoretical detection limits, we find that the system can provide ultra-sensitive single photon detection even in the sub-THz range. Owing to the ultra-low electronic heat capacity of the bilayer graphene, which is 2-3 orders of magnitude smaller than in any other superconducting photodetector, we find a remarkably fast response time of ~4 ns as well as an ultra-sensitive energy resolution better than 1 THz. The system can provide high-sensitivity single photon detection, which is indispensable for a broad range of research areas including quantum information processing and the most challenging applications in radio astronomy.

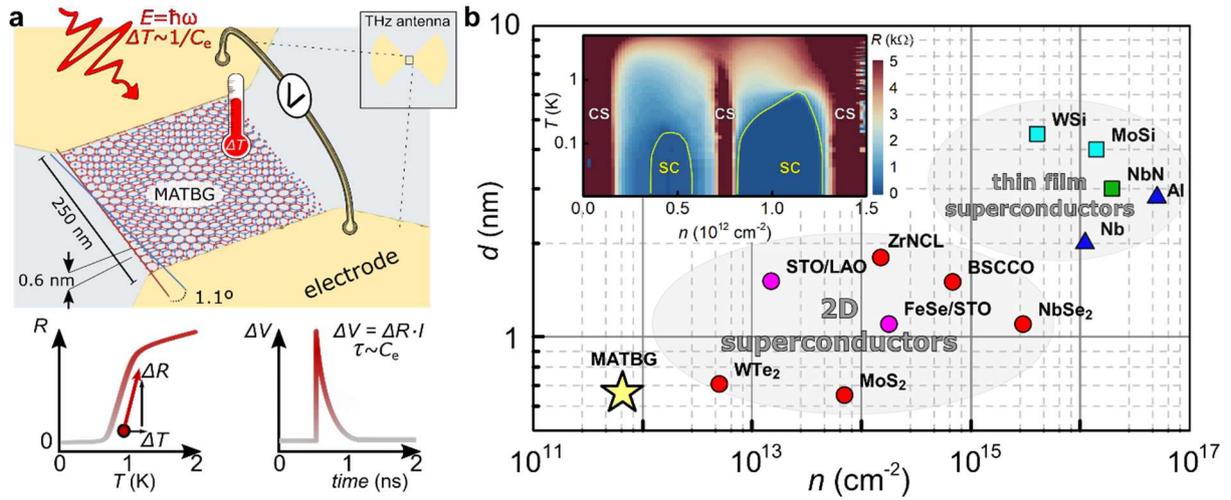

**Figure 1 | Magic-angle graphene superconducting single photon nano-calorimeter:**
**a** Schematic illustration of the MAG device. A twisted bilayer of graphene with twist angle of 1.1 degrees is sandwiched between two sheets of hBN and embedded in a lateral gold/MAG/gold photodetector geometry on a Si/SiO$_2$ substrate comprising a local graphite gate-electrode (not shown). At low temperatures the sheet resistance drops to zero at optimum doping. When a photon of a certain energy $E = \hbar\omega$ is absorbed the temperature of the MAG sheet is driven across the superconducting transition edge, giving rise to a voltage drop proportional to an applied bias current. The voltage response relaxes with the thermal relaxation time of the system. **b** Film thickness and carrier density for different two-dimensional superconductors and selected thin film superconductors below 10 nm which are commonly used in single photon detection applications. The plot includes crystalline 2D superconductors (red circles), interfacial 2D superconductors (pink circles), elemental thin film superconductors (blue triangles) and compound thin film superconductors of crystalline (green square) and amorphous (light green square) materials as well as MAG (yellow star). The inset depicts the experimentally obtained resistance of our MAG device as a function of carrier density $n$ and base temperature $T$, exhibiting a variety of phases including metallic, correlated (CS) and superconducting (SC) states. Data for carrier densities is extracted from Refs (*34–44*).

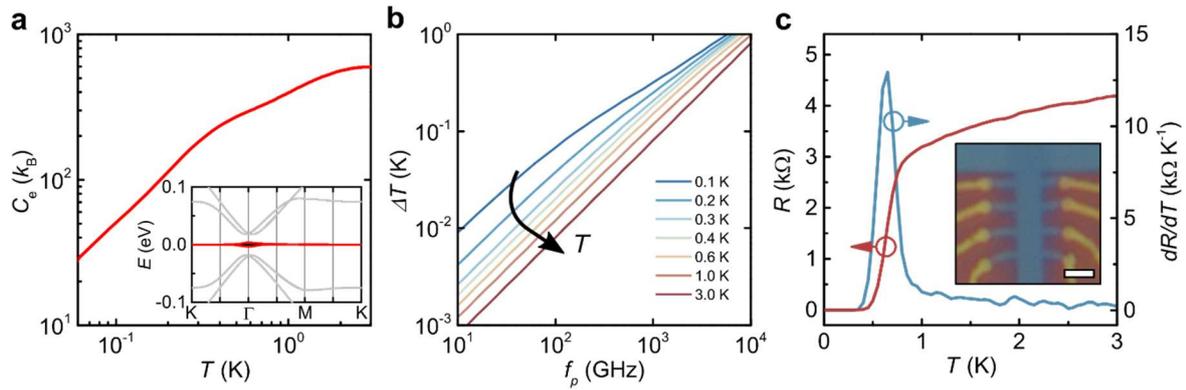

**Figure 2 | Intrinsic thermal properties of the MAG device:**
**a,** Electronic heat capacity as a function of $T$ as calculated from the density of states. The inset depicts the calculated low-energy band structure at the magic angle. **b,** Photon-absorption-induced temperature increase $\Delta T$ of the superconducting dome at $n = 1.1 \cdot 10^{12}/\text{cm}^2$ as a function of the photon frequency for different base temperatures. **c,** Resistance $R$ (red) and dR/dT (blue) of our MAG device at $n = 1.1 \cdot 10^{12}/\text{cm}^2$ as a function of temperature $T$. The inset depicts an optical microscopy image of the MAG device. Scale bar is 2 μm.

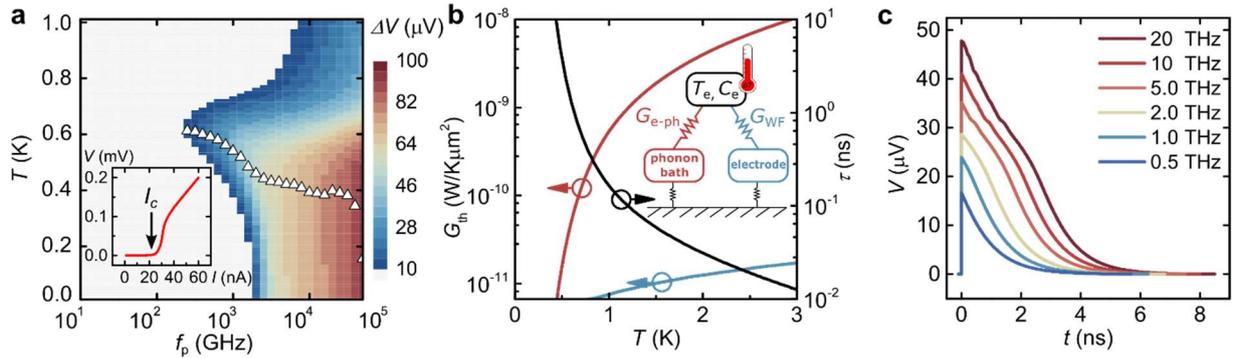

**Figure 3 | Photoresponse and relaxation time in magic-angle graphene:**
**a,** Photon-induced voltage drop $\Delta V$ across the MAG sheet at an applied bias current of 20 nA as a function of temperature and photon frequency. The white triangles indicate the points of maximum voltage response. The inset shows the experimentally obtained $I/V$ characteristics of the MAG device. **b,** Thermal conductance contributed by heat dissipation via electron–phonon scattering $G_{e\text{-}ph}$ (red) and the Wiedemann Frantz law $G_{WF}$ (blue), and thermal relaxation time $\tau_{th}$ (black) as a function of the device temperature. The inset depicts an illustration of heat dissipation channels of hot electrons in MAG after photon absorption. **c,** Transient voltage response of the MAG sheet at $T = T_c$ due to the absorption of a photon for frequencies from 0.5 THz to 20 THz

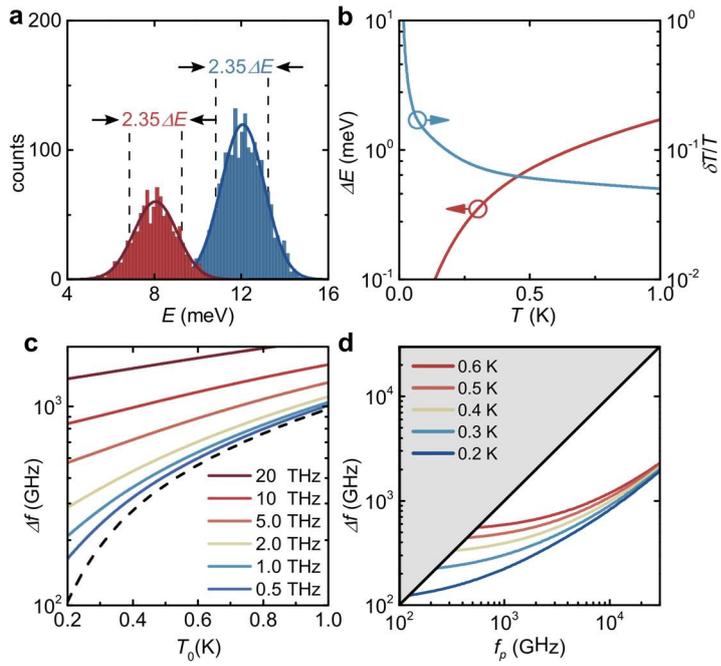

**Figure 4 | Thermodynamic fluctuations and energy resolution:**
**a**, Histogram of the potentially detected energy distribution of photons at two different energies after one thousand measurements. Due to the energetic broadening, two photon energies can be distinguished only if their energies are separated by more than the full width at half maximum of the distributions. **b**, Intrinsic energy scale $\Delta E$ of thermodynamic fluctuations (red) and associated relative temperature fluctuations $\delta T/T$ (blue) as calculated from the electronic heat capacity. **c**, Frequency resolution $\Delta f$ of single photon detection in MAG in the thermodynamic limit as a function of device temperature in the intrinsic case (black dashed line) and corrected for absorption of a photon for frequencies from 0.5 THz to 20 THz. **d,** Corresponding frequency resolution $\Delta f$ as a function of absorbed photon frequency for initial temperatures between 0.2 K and 0.6 K.

1. F. Marsili, V. B. Verma, J. A. Stern, S. Harrington, A. E. Lita, T. Gerrits, I. Vayshenker, B. Baek, M. D. Shaw, R. P. Mirin, S. W. Nam, Detecting single infrared photons with 93% system efficiency. *Nat. Photonics*. **7**, 210–214 (2013).
2. G. N. Gol'tsman, O. Okunev, G. Chulkova, A. Lipatov, A. Semenov, K. Smirnov, B. Voronov, A. Dzardanov, C. Williams, R. Sobolewski, Picosecond superconducting single-photon optical detector. *Appl. Phys. Lett.* **79**, 705–707 (2001).
3. Single Quantum Eos - SNSPD Closed-Cycle System (Single Quantum, 2019), (available at https://singlequantum.com/products/single-quantum-eos/).
4. D. H. Andrews, W. F. Brucksch, W. T. Ziegler, E. R. Blanchard, Attenuated Superconductors I. For Measuring Infra-Red Radiation. *Rev. Sci. Instrum.* **13**, 281–292 (1942).
5. K. D. Irwin, G. C. Hilton, in *Cryogenic Particle Detection, Topics in Applied Physics, vol 99* (Springer, Berlin, Heidelberg, Berlin, 2005; https://link.springer.com/chapter/10.1007%2F10933596_3), pp. 63–150.
6. M. D. Eisaman, J. Fan, A. Migdall, S. V. Polyakov, Single-photon sources and detectors. *Rev. Sci. Instrum.* **82**, 071101 (2011).
7. A. D. Semenov, G. N. Gol'tsman, A. A. Korneev, Quantum detection by current carrying superconducting film. *Phys. C Supercond.* **351**, 349–356 (2001).
8. V. B. Verma, A. E. Lita, B. A. Korzh, E. Wollman, M. Shaw, R. P. Mirin, S.-W. Nam, in *Proc. SPIE 10978, Advanced Photon Counting Techniques XIII*, M. A. Itzler, K. A. McIntosh, J. C. Bienfang, Eds. (SPIE, 2019; https://www.spiedigitallibrary.org/conference-proceedings-of-spie/10978/2519474/Towards-single-photon-spectroscopy-in-the-mid-infrared-using-superconducting/10.1117/12.2519474.full), p. 109780N.
9. K. D. Irwin, An application of electrothermal feedback for high resolution cryogenic particle detection. *Appl. Phys. Lett.* **66**, 1998–2000 (1995).
10. B. S. Karasik, S. V. Pereverzev, D. Olaya, M. E. Gershenson, R. Cantor, J. H. Kawamura, P. K. Day, B. Bumble, H. G. LeDuc, S. P. Monacos, D. G. Harding, D. Santavicca, F. Carter, D. E. Prober, in *Millimeter, Submillimeter, and Far-Infrared Detectors and Instrumentation for Astronomy V*, W. S. Holland, J. Zmuidzinas, Eds. (International Society for Optics and Photonics, 2010; http://proceedings.spiedigitallibrary.org/proceeding.aspx?doi=10.1117/12.856682), vol. 7741, p. 774119.
11. D. McCammon, in *Cryogenic Particle Detection, Topics in Applied Physics, vol 99* (Springer, Berlin, Heidelberg, 2005), pp. 1–34.
12. A. E. Lita, B. Calkins, L. A. Pellochoud, A. J. Miller, S. Nam, B. Young, B. Cabrera, A. Miller, in *AIP Conference Proceedings* (American Institute of Physics, 2009; http://aip.scitation.org/doi/abs/10.1063/1.3292350), vol. 1185, pp. 351–354.
13. D. Fukuda, G. Fujii, T. Numata, K. Amemiya, A. Yoshizawa, H. Tsuchida, H. Fujino, H. Ishii, T. Itatani, S. Inoue, T. Zama, Titanium-based transition-edge photon number resolving detector with 98% detection efficiency with index-matched small-gap fiber coupling. *Opt. Express*. **19**, 870 (2011).
14. R. Bistritzer, A. H. MacDonald, Moire bands in twisted double-layer graphene. *Proc. Natl. Acad. Sci.* **108**, 12233–12237 (2011).
15. Y. Cao, V. Fatemi, A. Demir, S. Fang, S. L. Tomarken, J. Y. Luo, J. D. Sanchez-Yamagishi, K. Watanabe, T. Taniguchi, E. Kaxiras, R. C. Ashoori, P. Jarillo-Herrero, Correlated insulator behaviour at half-filling in magic-angle graphene superlattices. *Nature*. **556**, 80–84 (2018).
16. Y. Cao, V. Fatemi, S. Fang, K. Watanabe, T. Taniguchi, E. Kaxiras, P. Jarillo-Herrero, Unconventional superconductivity in magic-angle graphene superlattices. *Nature*. **556**,

43–50 (2018).
17. X. Lu, P. Stepanov, W. Yang, M. Xie, M. A. Aamir, I. Das, C. Urgell, K. Watanabe, T. Taniguchi, G. Zhang, A. Bachtold, A. H. MacDonald, D. K. Efetov, Superconductors, Orbital Magnets, and Correlated States in Magic Angle Bilayer Graphene (2019) (available at http://arxiv.org/abs/1903.06513).
18. L. Ju, X. Xie, W.-C. Du, Y.-J. Liu, J.-J. Hao, B.-L. Ma, H.-W. Yang, Perfect Absorption in One-Dimensional Photonic Crystal with Graphene-Dielectric Hyperbolic Metamaterials. *Phys. status solidi*. **256**, 1800382 (2019).
19. D. K. Efetov, R.-J. Shiue, Y. Gao, B. Skinner, E. D. Walsh, H. Choi, J. Zheng, C. Tan, G. Grosso, C. Peng, J. Hone, K. C. Fong, D. Englund, Fast thermal relaxation in cavity-coupled graphene bolometers with a Johnson noise read-out. *Nat. Nanotechnol.* **13**, 797–801 (2018).
20. X. Gan, K. F. Mak, Y. Gao, Y. You, F. Hatami, J. Hone, T. F. Heinz, D. Englund, Strong Enhancement of Light–Matter Interaction in Graphene Coupled to a Photonic Crystal Nanocavity. *Nano Lett.* **12**, 5626–5631 (2012).
21. M. Furchi, A. Urich, A. Pospischil, G. Lilley, K. Unterrainer, H. Detz, P. Klang, A. M. Andrews, W. Schrenk, G. Strasser, T. Mueller, Microcavity-Integrated Graphene Photodetector. *Nano Lett.* **12**, 2773–2777 (2012).
22. J. Wang, Z. Cheng, C. Shu, H. K. Tsang, Optical Absorption in Graphene-on-Silicon Nitride Microring Resonators. *IEEE Photonics Technol. Lett.* **27**, 1765–1767 (2015).
23. S. Castilla, B. Terrés, M. Autore, L. Viti, J. Li, A. Y. Nikitin, I. Vangelidis, K. Watanabe, T. Taniguchi, E. Lidorikis, M. S. Vitiello, R. Hillenbrand, K.-J. Tielrooij, F. H. L. Koppens, Fast and Sensitive Terahertz Detection Using an Antenna-Integrated Graphene pn Junction. *Nano Lett.* **19**, 2765–2773 (2019).
24. K. J. Tielrooij, J. C. W. Song, S. A. Jensen, A. Centeno, A. Pesquera, A. Zurutuza Elorza, M. Bonn, L. S. Levitov, F. H. L. Koppens, Photoexcitation cascade and multiple hot-carrier generation in graphene. *Nat. Phys.* **9**, 248–252 (2013).
25. M. Koshino, N. F. Q. Yuan, T. Koretsune, M. Ochi, K. Kuroki, L. Fu, Maximally Localized Wannier Orbitals and the Extended Hubbard Model for Twisted Bilayer Graphene. *Phys. Rev. X*. **8**, 031087 (2018).
26. J. Wei, D. Olaya, B. S. Karasik, S. V. Pereverzev, A. V. Sergeev, M. E. Gershenson, Ultrasensitive hot-electron nanobolometers for terahertz astrophysics. *Nat. Nanotechnol.* **3**, 496–500 (2008).
27. K. C. Fong, E. E. Wollman, H. Ravi, W. Chen, A. A. Clerk, M. D. Shaw, H. G. Leduc, K. C. Schwab, Measurement of the electronic thermal conductance channels and heat capacity of graphene at low temperature. *Phys. Rev. X*. **3**, 041008 (2014).
28. E. D. Walsh, D. K. Efetov, G.-H. Lee, M. Heuck, J. Crossno, T. A. Ohki, P. Kim, D. Englund, K. C. Fong, Graphene-Based Josephson-Junction Single-Photon Detector. *Phys. Rev. Appl.* **8**, 024022 (2017).
29. M. Koshino, Y.-W. Son, Moiré phonons in the twisted bilayer graphene (2019) (available at http://arxiv.org/abs/1905.09660).
30. P. K. Day, H. G. LeDuc, B. A. Mazin, A. Vayonakis, J. Zmuidzinas, A broadband superconducting detector suitable for use in large arrays. *Nature*. **425**, 817–821 (2003).
31. T. C. P. Chui, D. R. Swanson, M. J. Adriaans, J. A. Nissen, J. A. Lipa, Temperature fluctuations in the canonical ensemble. *Phys. Rev. Lett.* **69**, 3005–3008 (1992).
32. A. G. Kozorezov, J. K. Wigmore, D. Martin, P. Verhoeve, A. Peacock, Resolution limitation in superconducting transition edge photon detectors due to downconversion phonon noise. *Appl. Phys. Lett.* **89** (2006), doi:10.1063/1.2397016.
33. C. B. McKitterick, D. E. Prober, B. S. Karasik, Performance of graphene thermal photon detectors. *J. Appl. Phys.* **113** (2013), doi:10.1063/1.4789360.

34. Y. Saito, T. Nojima, Y. Iwasa, Highly crystalline 2D superconductors. *Nat. Rev. Mater.* **2**, 16094 (2016).
35. X. Xi, H. Berger, L. Forró, J. Shan, K. F. Mak, Gate Tuning of Electronic Phase Transitions in Two-Dimensional $NbSe_2$. *Phys. Rev. Lett.* **117**, 106801 (2016).
36. W. Weingarten, Field-dependent surface resistance for superconducting niobium accelerating cavities. *Phys. Rev. Spec. Top. - Accel. Beams*. **14**, 101002 (2011).
37. N. Reyren, S. Thiel, A. D. Caviglia, L. F. Kourkoutis, G. Hammerl, C. Richter, C. W. Schneider, T. Kopp, A.-S. Rüetschi, D. Jaccard, M. Gabay, D. A. Muller, J.-M. Triscone, J. Mannhart, Superconducting interfaces between insulating oxides. *Science (80-. )*. **317**, 1196–9 (2007).
38. V. Fatemi, S. Wu, Y. Cao, L. Bretheau, Q. D. Gibson, K. Watanabe, T. Taniguchi, R. J. Cava, P. Jarillo-Herrero, Electrically tunable low-density superconductivity in a monolayer topological insulator. *Science*. **362**, 926–929 (2018).
39. Y. Saito, Y. Kasahara, J. Ye, Y. Iwasa, T. Nojima, Metallic ground state in an ion-gated two-dimensional superconductor. *Science*. **350**, 409–13 (2015).
40. P. Townsend, S. Gregory, R. G. Taylor, Superconducting Behavior of Thin Films and Small Particles of Aluminum. *Phys. Rev. B*. **5**, 54–66 (1972).
41. B. Z. Li, R. G. Aitken, Electrical transport properties of tungsten silicide thin films. *Appl. Phys. Lett.* **46**, 401–403 (1985).
42. S. P. Chockalingam, M. Chand, J. Jesudasan, V. Tripathi, P. Raychaudhuri, Superconducting properties and Hall effect of epitaxial NbN thin films. *Phys. Rev. B*. **77**, 214503 (2008).
43. A. Banerjee, L. J. Baker, A. Doye, M. Nord, R. M. Heath, K. Erotokritou, D. Bosworth, Z. H. Barber, I. MacLaren, R. H. Hadfield, Characterisation of amorphous molybdenum silicide (MoSi) superconducting thin films and nanowires. *Supercond. Sci. Technol.* **30**, 084010 (2017).
44. Y. Sun, W. Zhang, Y. Xing, F. Li, Y. Zhao, Z. Xia, L. Wang, X. Ma, Q.-K. Xue, J. Wang, High temperature superconducting FeSe films on SrTiO3 substrates. *Sci. Rep.* **4**, 6040 (2015).

# Supporting information

**Supplementary note 1: Calculation of heat capacity and cooling time**

To determine heat capacity and cooling time, we start from the kinetic equation in the absence of external fields and particle flow for the distribution function of electrons with momentum $\mathbf{k}$ and in band $\lambda$, $f_{\mathbf{k},\lambda}$, i.e.(1)

$$\partial_t f_{\mathbf{k},\lambda} = I[f_{\mathbf{k},\lambda}], \quad (1)$$

where the collision integral of the electron-phonon interaction reads

$$I[f_{\mathbf{k},\lambda}] = 2\pi \sum_{\mathbf{k}',\lambda'} \sum_{\mathbf{q},\nu} |V(\mathbf{q},\nu)|^2 \{D_{\mathbf{k},\lambda;\mathbf{k}',\lambda'} [f_{\mathbf{k},\lambda}(1 - f_{\mathbf{k}',\lambda'})n_{\mathbf{q},\nu} - f_{\mathbf{k}',\lambda'}(1 - f_{\mathbf{k},\lambda})(n_{\mathbf{q},\nu} + 1)] \delta(\mathbf{k}' - \mathbf{k} - \mathbf{q}) \, \delta(\varepsilon_{\mathbf{k}',\lambda'} - \varepsilon_{\mathbf{k},\lambda} - \omega_{\mathbf{q},\nu}) + \{\mathbf{k}, \lambda \leftrightarrow \mathbf{k}', \lambda'\}\}. \quad (2)$$

Here $V(\mathbf{q}, \nu)$ is the interaction between electrons and the phonon mode $\nu$ (e.g., longitudinal or transverse), and $D_{\mathbf{k},\lambda;\mathbf{k}',\lambda'}$ is the modulus square of the matrix element between the initial and final states $\mathbf{k}, \lambda$ and $\mathbf{k}', \lambda'$ of the electronic operator to which the phonon displacement is coupled. For the sake of the definiteness, we will assume the operator to be the electronic density. Other phonon models have been addressed in the literature[1], but will not be discussed here where the focus is to provide an order-of-magnitude estimate for the cooling time. In Eq. (2), $\varepsilon_{\mathbf{k},\lambda}$ and $\omega_{\mathbf{q},\nu}$ are respectively the electron and phonon energies, while $f_{\mathbf{k},\lambda}$ and $n_{\mathbf{q},\nu}$ are their distribution functions. In equilibrium, $f_{\mathbf{k},\lambda}$ ($n_{\mathbf{q},\nu}$) is the Fermi-Dirac (Bose-Einstein) distribution.

We now assume that $f_{\mathbf{k},\lambda}$ and $n_{\mathbf{q},\nu}$ are the Fermi-Dirac and Bose-Einstein distribution at the temperature $T_e$ and $T_L$, respectively. Each of the two subsystems (electrons and lattice vibrations) are therefore in thermal equilibrium, but the system as a whole is not. To determine the rate of heat conduction between them, we now multiply Eq. (1) by $\varepsilon_{\mathbf{k},\lambda} - \mu$, where $\mu$ is the chemical potential, and we integrate it over $\mathbf{k}$ and sum over $\lambda$(1). Expanding for $T_e \to T_L$, we get(1) $C \, \partial_t T_e = \Sigma \, (T_e - T_L)$, (3)
where

$$C = \sum_{\mathbf{k},\lambda} (\varepsilon_{\mathbf{k},\lambda} - \mu) \left(-\frac{\partial f_{\mathbf{k},\lambda}}{\partial \varepsilon_{\mathbf{k},\lambda}}\right) \left(\frac{\varepsilon_{\mathbf{k},\lambda} - \mu}{T_e} + \frac{\partial \mu}{\partial T_e}\right) \quad (4)$$

is the heat capacity, and

$$\Sigma = -2 \sum_{\mathbf{q},\nu} |V(\mathbf{q},\nu)|^2 \frac{\omega_{\mathbf{q},\nu}^2}{T_L} \left(-\frac{\partial n_{\mathbf{q},\nu}}{\partial \omega_{\mathbf{q},\nu}}\right) \text{Im}[\chi(\mathbf{q}, \omega_{\mathbf{q},\nu})]. \quad (5)$$

In Eq. (4),(1)

$$\frac{\partial \mu}{\partial T_e} = -\frac{\sum_{\mathbf{k},\lambda} \left(-\frac{\partial f_{\mathbf{k},\lambda}}{\partial \varepsilon_{\mathbf{k},\lambda}}\right) \left(\frac{\varepsilon_{\mathbf{k},\lambda} - \mu}{T_e}\right)}{\sum_{\mathbf{k},\lambda} \left(-\frac{\partial f_{\mathbf{k},\lambda}}{\partial \varepsilon_{\mathbf{k},\lambda}}\right)}, \quad (6)$$

which is obtained by assuming the density to be independent of $T_e$ (and fixed, e.g., by an external gate). In Eq. (5),

$$\text{Im}[\chi(\mathbf{q}, \omega_{\mathbf{q},\nu})] = -\pi \sum_{\mathbf{k},\lambda,\lambda'} (f_{\mathbf{k},\lambda} - f_{\mathbf{k}+\mathbf{q},\lambda'}) D_{\mathbf{k},\lambda;\mathbf{k}+\mathbf{q},\lambda'} \delta(\varepsilon_{\mathbf{k}+\mathbf{q},\lambda'} - \varepsilon_{\mathbf{k},\lambda} - \omega_{\mathbf{q},\nu}). \quad (7)$$

The cooling time is therefore defined by $\tau^{-1} = \Sigma/C$.

We now provide an estimate for the cooling time. Our goal is to approximate $\text{Im}[\chi(\boldsymbol{q}, \omega_{q,\nu})]$ in Eq. (7). To do so, we note that at $T \sim 1$ K only phonons with energies of the order of $4 k_B T \sim 0.3$ meV contribute to the integral, thanks to the derivative of the Bose-Einstein distribution which strongly suppresses higher energy excitations. Such energies correspond to phonon momenta of the order of $q \sim 0.05$ nm$^{-1}$ (using a phonon velocity $c_{ph} = 10^4$ m/s). Typical electron momenta are of the order of $2\pi/L_{moire} \sim 0.1 - 0.4$ nm$^{-1}$, *i.e.* much larger than phonon momenta. We can therefore estimate $\text{Im}[\chi(\boldsymbol{q}, \omega_{q,\nu})]$ in the limit of zero temperature and $q \to 0$. Restricting to the two flat bands, and approximating the matrix element as $D_{k,\lambda;k',\lambda'} = 1$ (which provides us with an upper limit to $\tau^{-1}$), and assuming that the bands are nearly particle-hole symmetric, after a few manipulations we get

$$\text{Im}[\chi(\boldsymbol{q}, \omega_{q,\nu})] \cong \frac{\pi}{2}\left[f\left(\frac{\omega - 2\mu}{2k_B T_e}\right) - f\left(-\frac{\omega + 2\mu}{2k_B T_e}\right)\right] N\left(\frac{\omega}{2}\right), \quad (7)$$

where $f(x) = (e^x + 1)^{-1}$ is the Fermi-Dirac distribution and $N(\varepsilon)$ is the density of states at the energy $\varepsilon$. $N(\varepsilon)$ is calculated from the continuum model of Ref. (*2*). Eq. (5) is then readily evaluated with (*3*)

$$|V(\boldsymbol{q}, \nu)|^2 = \frac{\hbar\, g^2 q}{2\, \rho\, c_{ph}}, \quad (8)$$

where(*3, 4*) $g = 3.6$ eV, $\rho = 7.6 \times 10^{-7}$ kg/m$^2$ and $\hbar$ is the reduced Planck's constant. Note that, by knowing the expression for the density of states $N(\varepsilon)$, the integrals over momenta in Eqs. (4) and (6) can be readily recast into integrals over band energies.

**Supplementary note 2: Calculation of detector response and energy resolution:**

After calculating the electronic heat capacity $C_e(T)$ for MATBG as a function of temperature (supplementary note 1), we calculate the photon-induced temperature increase by equating the energy of an absorbed photon $E_{\text{photon}}$ with the temperature-induced increase in internal energy

$$E_{\text{photon}} = h \cdot f_{\text{photon}} = \int_{T_0}^{T_{\text{max}}} C_e(T) dT. \quad (9)$$

Here $h$ is Planck's constant, $f_{\text{photon}}$ is the frequency of the absorbed photon, $T_0$ is the temperature of the MATBG before photon absorption and $T_{\text{max}}$ is the temperature of the MATBG directly after photon absorption. Solving for $T_{\text{max}}$ as a function of $f_{\text{photon}}$ and $T_0$ allows us to calculate the photon energy-dependent thermal response of the MATBG sheet.
With the experimentally obtained $R(T)$ and the calculated temperature increase $\Delta T(T_0, f_{\text{photon}}) = T_{\text{max}}(f_{\text{photon}}, T_0) - T_0$, we calculate the change in the MATBG's resistance upon absorption of a photon. Using a current $I$ just below the experimentally obtained critical current $I_c$, we calculate the voltage drop $\Delta V(T_0, f_{\text{photon}}) = I \cdot \Delta R(T_0, f_{\text{photon}})$.
Due to the fast ~100-fs thermalization time(*5*) after photon absorption, we assume the rise-time of the temperature transient to be instantaneous compared to the subsequent thermal relaxation, which we model by an exponential decay with time constant $\tau$ as obtained from the calculations in supplementary note 1. The corresponding transient voltage response is then calculated from $R(T(t))$.
On a timescale of the system's thermal time constant, the internal energy of a calorimeter in thermal equilibrium with the bath fluctuates by an amount $\langle \Delta E^2 \rangle = k_B T^2 C$ (*6*). This energy scale determines the uncertainty of any given energy measurement in a calorimeter and is such regarded as the thermodynamic limit on the energy resolution of the calorimeter. We take the full width at half maximum of the distribution in $E$ as the energetic discrimination threshold to distinguish the energies of two incident photons.

**Supporting figures:**

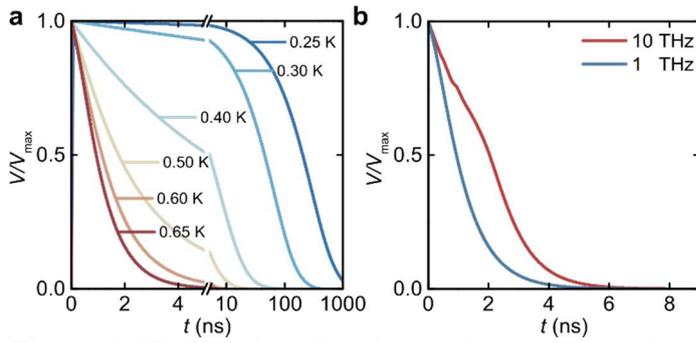

**Figure 1 | Relaxation time in magic-angle graphene:**
**a,** Relative voltage relaxation after the absorption of a 1-THz photon for different device temperatures. **b,** Relative voltage relaxation after the absorption of a photon of 1 THz (red) and 10 THz (blue) for $T=T_c$.

Figure 1 (a) shows the transient thermal voltage response of the device after photon absorption at a for photon frequency of 1 THz for temperatures between 0.25 K and 0.65 K. Remarkably, at $T = T_c$ the hot electron distribution relaxes within ∼4 ns for all photon frequencies. This is orders of magnitude faster than other energy-resolving superconducting single photon detectors, which exhibit recovery times on the order of tens of microseconds. At lower device temperatures the decreasing electron-phonon interaction leads to strongly increasing relaxation times surpassing 100 ns at $T = 0.3$ K and even 1 µs at $T = 0.25$ K. Figure 1 (b) shows the relative voltage relaxation after the absorption of a photon of 1 THz (red) and 10 THz (blue) for T =Tc. Due to the high temperature increase and the strongly non–linear dependence of the sample resistance on the device temperature, the relative voltage relaxation is slower for increasing photon frequency.

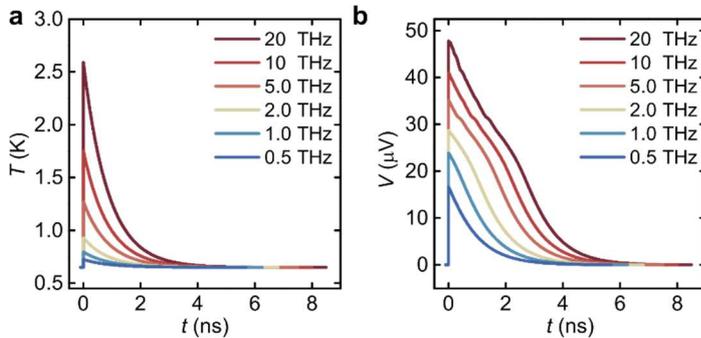

**Figure 2 | Transient relaxation of voltage response:**
**a,** Transient thermal response of the MATBG sheet at $T=T_c$ due to the absorption of a photon for frequencies from 0.5 THz to 20 THz. **b** Corresponding transient voltage response according to the temperature transient in (a).

Figure 2 (a and b) shows a comparison between the transient thermal response and transient voltage response of the device after photon absorption at a temperature of $T=T_c$ for photon frequencies between 0.5 THz and 20 THz. While the relative temperature relaxation is almost unchanged for increasing photon frequency, the relative voltage response deviates for higher photon frequencies due to the higher temperature increase and the strongly non–linear dependence of the sample resistance on the device temperature.

1. A. Principi, M. B. Lundeberg, N. C. H. Hesp, K. J. Tielrooij, F. H. L. Koppens, M. Polini, Super-Planckian Electron Cooling in a van der Waals Stack. *Phys. Rev. Lett.* **118**, 126804 (2017).
2. M. Koshino, N. F. Q. Yuan, T. Koretsune, M. Ochi, K. Kuroki, L. Fu, Maximally Localized Wannier Orbitals and the Extended Hubbard Model for Twisted Bilayer Graphene. *Phys. Rev. X.* **8**, 031087 (2018).
3. S. Das Sarma, S. Adam, E. H. Hwang, E. Rossi, Electronic transport in two-dimensional graphene. *Rev. Mod. Phys.* **83**, 407–470 (2011).
4. G. X. Ni, A. S. McLeod, Z. Sun, L. Wang, L. Xiong, K. W. Post, S. S. Sunku, B. Y. Jiang, J. Hone, C. R. Dean, M. M. Fogler, D. N. Basov, Fundamental limits to graphene plasmonics. *Nature.* **557**, 530–533 (2018).
5. K. J. Tielrooij, J. C. W. Song, S. A. Jensen, A. Centeno, A. Pesquera, A. Zurutuza Elorza, M. Bonn, L. S. Levitov, F. H. L. Koppens, Photoexcitation cascade and multiple hot-carrier generation in graphene. *Nat. Phys.* **9**, 248–252 (2013).
6. T. C. P. Chui, D. R. Swanson, M. J. Adriaans, J. A. Nissen, J. A. Lipa, Temperature fluctuations in the canonical ensemble. *Phys. Rev. Lett.* **69**, 3005–3008 (1992).